\documentclass[10pt, reprint, onecolumn, amssymb, amsmath, tightenlines, floatfix, citeautoscript, amsmath, aip]{revtex4-1}

\usepackage{fancyhdr}
\fancyheadoffset{0cm}
\usepackage{bm}
\usepackage{amsmath,mathtools}
\usepackage{amsfonts}
\usepackage{amssymb}
\usepackage{graphicx}
\usepackage{verbatim}
\usepackage{hyperref}
\usepackage{eucal}

\def\rcur{{\mbox{$\resizebox{.14in}{.08in}{\includegraphics{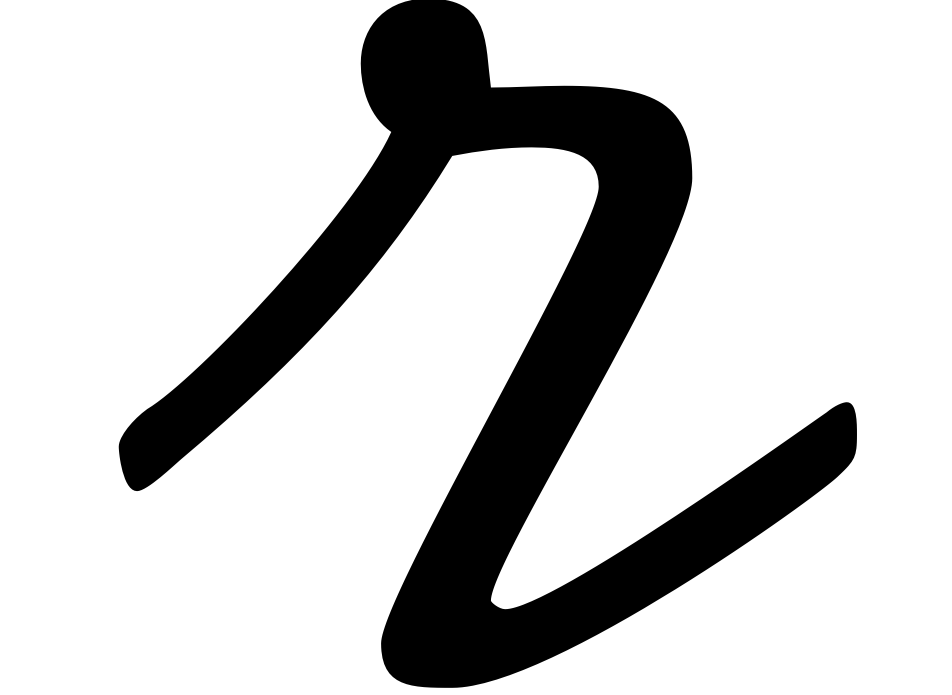}}$}}}

\newcommand{\hrcur}{\hat{\rcur}}

\newcommand{\D}{{\mathbb{D}}}

\newcommand{\R}{{\bf r}}

\newcommand{\K}{{\bf k}}

\newcommand{\phib}{\bar{\phi}}

\newcommand{\be}{\begin}
\newcommand{\en}{\end}
\newcommand{\eq}{equation}

\newcommand{\KK}{\mathbb{K}}

\begin{document}

\lhead{\small{D. M. Riffe, et al.}}
\rhead{\small{EAM -- BvK Vibrational Dynamics}}
\cfoot{--\thepage--}

\title{Vibrational Dynamics within the Embedded-Atom-Method Formalism and the Relationship to Born-von-K\'{a}rm\'{a}n Force Constants}


\author{D. M. Riffe}
\email[Author to whom correspondence should be addressed; electronic mail: ]{mark.riffe@usu.edu}
\author{Jake D. Christensen}

\affiliation{Physics Department, Utah State University, Logan, UT  84322-4415}

\author{R. B. Wilson}

\affiliation{Mechanical Engineering Department, University of California, Riverside, CA  92521}

\date{\today}

\begin{abstract}
We derive expressions for the dynamical matrix of a crystalline solid with total potential energy described by an embedded-atom-method (EAM) potential.  We make no assumptions regarding the number of atoms per unit cell.  These equations can be used for calculating both bulk phonon modes as well the modes of a slab of material, which is useful for the study of surface phonons.  We further discuss simplifications that occur in cubic lattices with one atom per unit cell.  The relationship of Born-von-K\'{a}rm\'{a}n (BvK) force constants  -- which are readily extracted from experimental vibrational dispersion curves -- to the EAM potential energy is discussed.  In particular, we derive equations for BvK force constants for bcc and fcc lattices in terms of the functions that define an EAM model.  The EAM -- BvK relationship is useful for assessing the suitability of a particular EAM potential for describing vibrational spectra, which we illustrate using vibrational data from the bcc metals K and Fe and the fcc metal Au.

\end{abstract}


\maketitle

\section{Introduction}

Embedded-atom-method (EAM) models are popularly used to calculate vibrational properties of crystalline materials, both in the bulk and at surfaces \cite{*[{See, e.g., }] [] Rusina2013}.  The key to these calculations is a quantity known as the dynamical matrix $\mathbb{D}$:  the eigenvalues and eigenvectors of $\mathbb{D}$ respectively give the normal-mode frequencies and polarizations.  However, (i) the few equations for $\mathbb{D}$ that appear in the literature are only applicable to solids with one atom per unit cell, and (ii) discrepancies exist among these equations for $\D$ \cite{DAWssc1985, NINGSHENGssc1989, WBJAP1995, KAZANCphysicab2005, KAZANCphysicab2006}.

To the end of having an accurate set of expressions for $\mathbb{D}$ that can be used for any crystalline solid, here we derive general equations for $\mathbb{D}$ within the EAM formalism.  These general expressions can be used for finding bulk vibrational modes in monatomic materials as well as crystalline alloys.  Vibrational modes of a slab -- which is the typical setting for studying vibrations at the surface of a solid -- can also be investigated using our derived equations.

EAM modeling of vibrations is perhaps most commonly used to study bcc and fcc materials; we therefore also derive simplified expressions that are applicable to these materials.  Furthermore, as vibrational data from materials with these two lattice structures are often analyzed to extract Born-von-K\'{a}rm\'{a}n (BvK) atomic force constants (FCs), we derive equations that relate the BvK constants to a general EAM potential.  Using K, Fe, and Au as examples, we conclude by illustrating how BvK analysis gives insight into the suitability of a particular EAM potential for describing vibrational properties.

\section{EAM Model}

In this section we outline the standard embedded-atom-method model \cite{DawPRB1984}.  Our presentation is general enough that it can be can be applied to materials with multiple types of atoms in each unit cell.

The total energy $E$ of a solid in the EAM formalism is written as
\be{\eq} 
\label{e1}
E = E^{p} + E^{e},
\en{\eq}
where
\be{\eq} 
\label{e2}
E^{p} = \frac{1}{2} \sum_{n\alpha} \sum_{m\beta} \phi_{\alpha}^{\beta}(\rcur_{n\alpha}^{m\beta})
\en{\eq}
and
\be{\eq} 
\label{e3}
E^{e} = \sum_{n\alpha} F_{\alpha}(\rho(\R_{n\alpha})).
\en{\eq}
Here $E^{p}$ is a sum of interatomic pair potentials $\phi_{\alpha}^{\beta} (= \phi_{\beta}^{\alpha})$, where $m$ and $n$ label the unit cells of the solid and $\alpha$ and $\beta$ label the (perhaps different types of) atoms within each unit cell.  The combination $m\alpha$ (for example) thus accounts for all atoms in the solid.  The argument $\rcur_{n\alpha}^{m\beta}$ of the pair potential is the distance between the atoms labeled by $m\alpha$ and $n\beta$.  That is, $\rcur_{n\alpha}^{m\beta} = | \R_{m\beta} - \R_{n\alpha} | $.  The potential $\phi_{\alpha}^{\beta}$ is thus associated with a central force.  In the sums in (\ref{e2}) terms with $n\alpha = m\beta$ are excluded as these would correspond to a self-interaction.  The energy $E^{e}$ is the sum of individual energies $F_{\alpha}$ associated with embedding each atom in a background charge density $\rho$ at that atom's position $\R_{n\alpha}$.  As is standard practice, we assume $\rho(\R_{n\alpha})$ to be a sum of individual atomic charge densities,
\be{\eq} 
\label{e4}
\rho(\R_{n\alpha}) = \sum_{m\beta} f_{\beta}(\rcur^{n\alpha}_{m\beta}).
\en{\eq}
Here $f_{\beta}(\rcur^{n\alpha}_{m\beta})$ is the charge density from atom $m\beta$ at the location of atom $n\alpha$.  That $f_{\beta}$ is a function of $\rcur^{n\alpha}_{m\beta}$ follows from the assumption the atomic charge densities have spherical symmetry.  

The indices $\alpha$ and/or $\beta$ on $\phi$, $F$, and $f$ are the minimum number of required indices, as these functions are not expected to be the same for different types of atoms.  However, in the interest of notational simplicity, we enlarge the number of  indices by making the definitions
\be{\eq} 
\label{e4a}
\phi^{m\beta}_{n\alpha} = \phi^{\beta}_{\alpha}(\rcur^{m\beta}_{n\alpha}), 
\en{\eq}
\be{\eq} 
\label{e4b}
F_{n\alpha} = F_\alpha (\rho(\R_{n\alpha})),
\en{\eq}
and
\be{\eq} 
\label{e4c}
f_{m\beta}^{n\alpha} = f_{\beta}(\rcur_{m\beta}^{n\alpha}).
\en{\eq}
These functions define any EAM model; as we see below, derivatives of these functions are key components of the dynamical matrix.

We note an extension to embedded-atom method presented here, known as the modified analytic EAM (MAEAM), has been used to calculate vibrational properties of alkali and noble metals \cite{HuMSMSE2002, ZhangJLTP2008, XieCJP2008, Xie2008, Ram2016}.  Originally introduced to account for the negative Cauchy pressure in Cr, the key feature of the MAEAM is an additional energy term that depends upon the square of the atomic charge density \cite{Yifang1996}.  However, as the MAEAM does not appear to possess any advantage for describing vibrational structure, we shall not consider it further.

\section{Dynamical Matrix}

Here we outline a description of the vibrational dynamics of a solid with energy $E$ as given above.  Our overall goal is to find the normal modes of vibration of the atoms in the solid.  As we shall see, a quantity known as the dynamical matrix is the key to finding the frequencies and atomic displacements associated with these modes. 
 
\subsection{General Formalism}

We start with the general development of $\mathbb{D}$, which closely follows that of Ibach and L\"{u}th \cite{Ibach}.  If we assume the atoms in a crystal do not move far from their equilibrium positions, then we may expand the energy $E$ in a Taylor series in the set of displacements $\{{\bf s}_{n\alpha}\}$ of the atoms.  Up to second order the energy can be expressed as
\be{\eq}
\label{e4e}
E  = E_0 + \frac{1}{2} \sum_{n\alpha i} \sum_{m\beta j} K_{n\alpha i}^{m\beta j} \, s_{n\alpha i} \, s_{m\beta j}.
\en{\eq}
where $E_0$ is the equilibrium energy of the solid and 
\be{\eq} 
\label{e6}
K_{n\alpha i}^{m\beta j} = \frac{\partial^2 E}{\partial r_{n\alpha i} \partial r_{m\beta j}}.
\en{\eq}
Here $i$ and $j$ represent Cartesian coordinates.  We note the right side of (\ref{e6}) is evaluated at equilibrium, and the case $n\alpha = m\beta$ is not excluded.  

Utilizing (\ref{e4e}) Newton's second law gives us the equations of motion for the atoms,
\be{\eq}
\label{e6a}
M_\alpha \ddot{s}_{n\alpha i} = \sum_{m\beta j} \big( \! - K_{n\alpha i}^{m\beta j} \big)  \, s_{m\beta j},
\en{\eq}
where $M_\alpha$ is the mass of atoms labeled by $\alpha$.  This last equation tells us $- K_{n\alpha i}^{m\beta j}$ is the force on atom $n\alpha$ in the Cartesian direction $i$ when atom $m\beta$ is displaced in the Cartesian direction $j$ a unit distance.

If we now assume the displacements of all atoms are coherently related by a plane wave with wave vector $\K$ and angular frequency $\omega$,
\be{\eq}
\label{e6b}
s_{n\alpha i} = \frac{1}{\sqrt{M_\alpha}} \, u_{\alpha i} \, e^{i (\K \cdot \R_n - \omega t)},
\en{\eq} 
then the equations of motion for the atoms become a set of coupled algebraic equations,
\be{\eq}
\label{e6c}
\sum_{\beta j} \mathbb{D}_{\alpha i}^{\beta j}(\K) \, u_{\beta j} = \omega^2 \, u_{\alpha i},
\en{\eq}
where 
\be{\eq} 
\label{e5}
\mathbb{D}_{\alpha i}^{\beta j}(\K)  = \frac{1}{\sqrt{M_\alpha M_\beta}} \sum_m K_{n\alpha i}^{m\beta j} \, e^{i \K \cdot (\R_{m} - \R_{n})},
\en{\eq}
are the Cartesian components of a quantity known as the dynamical matrix $\mathbb{D}(\K)$.  Here $\R_{m} - \R_{n}$ in (\ref{e5}) is the displacement between identical locations within the $m$ and $n$ unit cells.  If there are $N$ atoms per unit cell, then $\D(\K)$ is a $3N \times 3N$ matrix.  The eigenvalues $\omega^2$ and eigenvectors $u_{\alpha i}$ of $\mathbb{D}$ characterize the normal modes of motion; for each wave vector $\K$ there are $3N$ vibrational modes.

\subsection{EAM Force Constants}

In the following two subsections we sequentially find the pair-potential contribution $\prescript{p}{}\D(\K)$ and the embedding-energy contribution $\prescript{e}{}\D(\K)$ to the dynamical matrix $\D(\K)= \prescript{p}{}\D(\K) + \prescript{e}{}\D(\K)$ for cases that include more than one atom per unit cell.  
\subsubsection{Pair-Potential Contribution}

For the pair-potential contribution $\prescript{p}{}{\D}(\K)$ to the dynamical matrix we require
\be{\eq} 
\label{e7}
\prescript{p \!}{}{K_{n\alpha i}^{m\beta j}} = \frac{\partial^2 E^{p}}{\partial r_{n\alpha i} \partial r_{m\beta j}}.
\en{\eq}
Using (\ref{e2}) this expression can be written as
\be{\eq} 
\label{e8}
\prescript{p \!}{}{K_{n\alpha i}^{m\beta j}} = \frac{1}{2}  \frac{\partial}{\partial r_{n\alpha i}} \Bigg[ \frac{\partial}{\partial r_{m\beta j}}  \sum_{n'\alpha'} \sum_{m'\beta'} \phi_{n'\alpha'}^{m'\beta'} \Bigg].
\en{\eq}
Within the double sum there is one set of terms with $n'\alpha' = m\beta$ and one set with $m'\beta' = m\beta$.  This last equation thus simplifies to
\be{\eq} 
\label{e9}
\prescript{p \!}{}{K_{n\alpha i}^{m\beta j}} = \frac{1}{2}  \frac{\partial}{\partial r_{n\alpha i}} \Bigg[  \sum_{m'\beta'} \frac{\partial \phi_{m\beta}^{m'\beta'}}{\partial r_{m\beta j}}  + \sum_{n'\alpha'} \frac{\partial \phi_{n'\alpha'}^{m\beta}}{\partial r_{m\beta j}}  \Bigg] .
\en{\eq}
Because (i) $\phi_{n\alpha}^{m\beta} = \phi_{m\beta}^{n\alpha}$ and (ii) the indices on the sums are dummy indices, this expression itself simplifies to
\be{\eq} 
\label{e10}
\prescript{p \!}{}{K_{n\alpha i}^{m\beta j}} =  \frac{\partial}{\partial r_{n\alpha i}}  \sum_{m'\beta'} \frac{\partial \phi_{m'\beta'}^{m\beta}}{\partial r_{m\beta j}}.
\en{\eq}
The only nonzero terms on the right side of this equation are those with either $m\beta = n\alpha$ or $m'\beta' = n\alpha$.  Therefore we can express (\ref{e10}) as
\be{\eq} 
\label{e11}
\prescript{p \!}{}{K_{n\alpha i}^{m\beta j}} = \sum_{m'\beta'} \frac{\partial^2 \phi_{m'\beta'}^{m\beta}}{ \partial r_{m\beta i} \partial r_{m\beta j}} \, \delta_{nm} \delta_{\alpha \beta} + \frac{\partial^2 \phi^{m\beta}_{n\alpha}}{\partial r_{n\alpha i} r_{m\beta j}} \, (1 - \delta_{nm} \delta_{\alpha \beta}),
\en{\eq}
where $\delta_{nm}$ is the standard Kronecker delta.  We note the last term on the right side of this equation is the FC associated with the force on atom $n\alpha$ when a different atom $m\beta$ is displaced, while the sum of terms (over $m'\beta'$) is the FC associated with the force on atom $m\beta$ when that same atom is displaced. 

We can make progress towards evaluating the derivatives in (\ref{e11}) owing to the pair potential $\phi^{m\beta}_{n\alpha}$ being related to $r_{m \beta j}$ and $r_{n \alpha, i}$ through its argument via
\be{\eq} 
\label{e12}
\rcur_{n\alpha}^{m\beta} = | \R_{m\beta} - \R_{n\alpha} | = \sqrt{(r_{m\beta x} - r_{n\alpha x})^2 + (r_{m\beta y} - r_{n\alpha y})^2 + (r_{m\beta z} - r_{n\alpha z})^2}
\en{\eq}
(where $i$ and $j$ are one of $x$, $y$, or $z$).  We require derivatives of this argument; from (\ref{e12}) we find
\be{\eq} 
\label{e13}
\frac{\partial \rcur_{n\alpha}^{m\beta} }{\partial r_{m\beta j}} = \frac{r_{m\beta j} - r_{n\alpha j}}{| \R_{m\beta} - \R_{n\alpha} |} = \hrcur_{n\alpha j}^{m\beta}
\en{\eq}
and
\be{\eq} 
\label{e14}
\frac{\partial \rcur_{n\alpha}^{m\beta} }{\partial r_{n\alpha j}} = \frac{ r_{n\alpha j} - r_{m\beta j} }{| \R_{m\beta} - \R_{n\alpha} |} = - \hrcur_{n\alpha j}^{m\beta},
\en{\eq}
where we have defined $\hrcur_{n\alpha j}^{m\beta}$ to be the $j$th component of the unit vector that points from atom $n\alpha$ to atom $m\beta$.  We also require derivatives of this unit vector.  Utilizing (\ref{e12}) -- (\ref{e14}) it is straightforward to verify
\be{\eq} 
\label{e15}
\frac{\partial \hrcur_{n\alpha j}^{m\beta}}{\partial r_{m\beta i}} = \frac{\partial }{\partial r_{m\beta i}} \frac{r_{m\beta j} - r_{n\alpha j}}{| \R_{m\beta} - \R_{n\alpha} |} = \frac{\delta_{ij}}{| \R_{m\beta} - \R_{n\alpha} |} - \frac{(r_{m\beta j} - r_{n\alpha j})(r_{m\beta i} - r_{n\alpha i})}{| \R_{m\beta} - \R_{n\alpha} |^3},
\en{\eq}
which we succinctly express as
\be{\eq} 
\label{e16}
\frac{\partial \hrcur_{n\alpha j}^{m\beta}}{\partial r_{m\beta i}}  = \frac{1}{\rcur_{n\alpha}^{m\beta}}  \big( \delta_{ij} - \hrcur_{n\alpha j}^{m\beta} \, \hrcur_{n\alpha i}^{m\beta}  \big).
\en{\eq}
Similarly, we find
\be{\eq} 
\label{e17}
\frac{\partial \hrcur_{n\alpha j}^{m\beta}}{\partial r_{n\alpha i}}  = - \frac{1}{\rcur^{m\beta}_{n\alpha}}  \big( \delta_{ij} - \hrcur_{n\alpha j}^{m\beta} \, \hrcur_{n\alpha i}^{m\beta}  \big).
\en{\eq}
Appealing to (\ref{e13}), (\ref{e14}), (\ref{e16}), and (\ref{e17}) we apply the chain rule to (\ref{e11}) to express the FC as
\begin{align}
\label{e18}
\prescript{p \!}{}{K_{n\alpha i}^{m\beta j}} =  &\sum_{m'\beta'}  \Bigg[  \frac{D\phi^{m\beta}_{m'\beta'}}{\rcur^{m\beta}_{m'\beta'}} \,  \Big( \delta_{ij} - \hrcur^{m\beta}_{m'\beta' i} \, \hrcur^{m\beta}_{m'\beta' j} \Big) +  D^2\phi^{m\beta}_{m'\beta'}  \hrcur^{m\beta}_{m'\beta' i} \, \hrcur^{m\beta}_{m'\beta' j} \Bigg] \, \delta_{nm} \delta_{\alpha \beta} \nonumber \\ 
& - \Bigg[  \frac{D\phi^{m\beta}_{n\alpha}}{\rcur^{m\beta}_{n\alpha}} \,  \Big( \delta_{ij} - \hrcur^{m\beta}_{n\alpha i} \, \hrcur^{m\beta}_{n\alpha j} \Big) +  D^2\phi^{m\beta}_{n\alpha}  \hrcur^{m\beta}_{n\alpha i} \, \hrcur^{m\beta}_{n\alpha j} \Bigg] \, (1 - \delta_{nm} \delta_{\alpha \beta}), 
\end{align}
where $D\phi_{n\alpha}^{m\beta}$ and $D^2\phi_{n\alpha}^{m\beta}$ are respectively the first and second derivatives of $\phi_{n\alpha}^{m\beta}$ with respect to its argument $\rcur_{n\alpha}^{m\beta}$ (evaluated at the equilibrium positions of the atoms).  When this expression is substituted into (\ref{e5}) for $K^{m\beta j}_{n\alpha i}$ one obtains the pair-potential part of the dynamical matrix $\prescript{p}{}{\D}^{\beta j}_{\alpha i}(\K)$.

\subsubsection{Embedding-Energy Contribution}

To find the embedding-energy contribution to the dynamical matrix we formally proceed as we just have for the pair-potential part.  The bulk of the calculation consists of finding an expression for the FCs 
\be{\eq} 
\label{e25}
\prescript{e \!}{}{K_{n\alpha i}^{m\beta j}} = \frac{\partial^2 E^{e}}{\partial r_{n\alpha i} \partial r_{m\beta j}}.
\en{\eq}
Using (\ref{e3}) we start by writing (\ref{e25}) as
\begin{equation}
\label{e26}
\prescript{e \!}{}{K_{n\alpha i}^{m\beta j}} = \frac{\partial}{\partial r_{n\alpha i}} \Bigg[ \sum_{n'\alpha'} \frac{\partial}{\partial r_{m\beta j}}  F_{\alpha'} \big(\Sigma_{m'\beta'} f_{\beta'}(\rcur^{n'\alpha'}_{m'\beta'})  \big)  \Bigg]. 
\end{equation}
Here we have also used (\ref{e4}) to explicitly express the argument $\rho(\R_{n'\alpha'})$ of $F_{\alpha'}$ in terms of the atomic charge densities $f_{\beta'}(\rcur^{n'\alpha'}_{m'\beta'})$.  As written, this equation portends the complexity of the final result.  From (\ref{e26}) we observe that nonzero terms only occur if $n'\alpha' = m\beta $ or $m'\beta' = m\beta $.  Applying the chain rule and using our previous expressions (\ref{e13}) and (\ref{e14}) for the derivatives of $\rcur_{n\alpha}^{m\beta}$ we straightforwardly obtain
\be{\eq}
\label{e27}
\prescript{e \!}{}{K_{n\alpha i}^{m\beta j}} = \frac{\partial}{\partial r_{n\alpha i}}   \sum_{m'\beta'} \Big( D\!F_{m\beta} D\!f^{m\beta}_{m'\beta'} + D\!F_{m'\beta'} D\!f^{m'\beta'}_{m\beta}  \Big) \hrcur^{m\beta}_{m'\beta'j},
\en{\eq}
where (as is the case with $D\phi_{n\alpha}^{m\beta}$), $D\!F_{n\alpha}$ and $D\!f_{n\alpha}^{m\beta}$ are derivatives of $F_{n\alpha}$ and $f_{n\alpha}^{m\beta}$ with respect to their arguments $\rho(\R_{n\alpha})$ and $\rcur_{n\alpha}^{m\beta}$, respectively (again, evaluated at the equilibrium positions of the atoms).  Quite obviously, we now need derivatives of three types of terms: $D\!f_{m'\beta'}^{m\beta}$, $D\!F_{m\beta}$, and $\hrcur^{m\beta}_{m'\beta' j}$.  Derivatives of the last of these three quantities are given by (\ref{e16}) and (\ref{e17}), while derivatives of the first two quantities can be expressed as
\be{\eq}
\label{e28}
\frac{\partial D\!f_{m'\beta'}^{m\beta}}{\partial r_{n\alpha i}} = D^2\!f_{m'\beta'}^{m\beta} \Big(    \hrcur^{m\beta}_{m'\beta' i} \, \delta_{nm}\delta_{\alpha \beta} -  \hrcur_{m'\beta' i}^{m\beta} \, \delta_{nm'}\delta_{\alpha \beta'}  \Big)
\en{\eq}
and
\be{\eq}
\label{e29}
\frac{\partial D\!F_{m\beta}}{\partial r_{n\alpha i}} = D^2\!F_{m\beta} \Bigg[ \sum_{m''\beta''} \Big( D\!f_{m''\beta''}^{m\beta} \hrcur_{m''\beta'' i}^{m\beta} \, \delta_{nm} \delta_{\alpha \beta} \Big)  - D\!f_{n\alpha}^{m\beta}  \hrcur_{n\alpha i}^{m\beta}\,  \big( 1 - \delta_{nm} \delta_{\alpha \beta} \big)   \Bigg].
\en{\eq}
Using (\ref{e16}), (\ref{e17}), (\ref{e28}), and (\ref{e29}) we rewrite (\ref{e27}) as
\be{\eq}
\label{e30}
\prescript{e \!}{}{K_{n\alpha i}^{m\beta j}} = \prescript{e1 \!}{}{K_{n\alpha i}^{m\beta j}} + \prescript{e2 \!}{}{K_{n\alpha i}^{m\beta j}} +\prescript{e3 \!}{}{K_{n\alpha i}^{m\beta j}},
\en{\eq}
where
\begin{align}
\label{e31}
\prescript{e1 \!}{}{K_{n\alpha i}^{m\beta j}} = &\sum_{m'\beta'} \Big( D\!F_{m\beta} D\!f^{m\beta}_{m'\beta'} + D\!F_{m'\beta'} D\!f^{m'\beta'}_{m\beta}  \Big) \frac{1}{\rcur^{m\beta}_{m'\beta'}}  \big( \delta_{ij} - \hrcur_{m'\beta' i}^{m\beta} \, \hrcur_{m'\beta' j}^{m\beta}  \big) \delta_{nm}\delta _{\alpha\beta} \nonumber\\ 
&- \Big( D\!F_{m\beta} D\!f^{m\beta}_{n\alpha} + D\!F_{n\alpha} D\!f^{n\alpha}_{m\beta}  \Big) \frac{1}{\rcur^{m\beta}_{n\alpha}}  \big( \delta_{ij} - \hrcur_{n\alpha i}^{m\beta} \, \hrcur_{n\alpha j}^{m\beta}  \big) \, (1 - \delta_{nm}\delta _{\alpha\beta}),
\end{align}
\begin{align}
\label{e32}
\prescript{e2 \!}{}{K_{n\alpha i}^{m\beta j}} = &\sum_{m'\beta'} \Big( D\!F_{m\beta} D^2\!f^{m\beta}_{m'\beta'} + D\!F_{m'\beta'} D^2\!f^{m'\beta'}_{m\beta}  \Big) \hrcur_{m'\beta' i}^{m\beta} \, \hrcur_{m'\beta' j}^{m\beta}  \, \delta_{nm}\delta _{\alpha\beta} \nonumber \\ 
&- \Big( D\!F_{m\beta} D^2\!f^{m\beta}_{n\alpha} + D\!F_{n\alpha} D^2\!f^{n\alpha}_{m\beta}  \Big)  \hrcur_{n\alpha i}^{m\beta} \, \hrcur_{n\alpha j}^{m\beta} \, (1 - \delta_{nm}\delta _{\alpha\beta}) ,
\end{align}
and
\begin{align}
\label{e33}
\prescript{e3 \!}{}{K_{n\alpha i}^{m\beta j}} = &\sum_{m'\beta'} \sum_{m''\beta''} \, D^2\!F_{m\beta} \,  D\!f^{m\beta}_{m''\beta''} \, D\!f^{m\beta}_{m'\beta'} \, \hrcur_{m''\beta'' i}^{m\beta} \, \hrcur_{m'\beta' j}^{m\beta}  \, \delta_{nm}\delta _{\alpha\beta} \nonumber \\ 
&-  \sum_{m'\beta'} D^2\!F_{m\beta} \,  D\!f^{m\beta}_{n\alpha} \, D\!f^{m\beta}_{m'\beta'} \, \hrcur_{n\alpha i}^{m\beta} \, \hrcur_{m'\beta' j}^{m\beta} \, (1 - \delta_{nm}\delta _{\alpha\beta}) \nonumber \\ 
&-  \sum_{m'\beta'} D^2\!F_{n\alpha} \,  D\!f^{n\alpha}_{m'\beta'} \, D\!f^{n\alpha}_{m\beta} \, \hrcur_{m'\beta' i}^{n\alpha} \, \hrcur_{m\beta j}^{n\alpha} \, (1 - \delta_{nm}\delta _{\alpha\beta}) \nonumber \\
& + \sum_{m'\beta'} D^2\!F_{m'\beta'} \,  D\!f_{n\alpha}^{m'\beta'} \, D\!f^{m'\beta'}_{m\beta} \, \hrcur_{n\alpha i}^{m'\beta'} \, \hrcur_{m\beta j}^{m'\beta'}.
\end{align}
We note the last term in (\ref{e33}) includes terms with $n\alpha = m\beta$ and $n\alpha \ne m\beta$.  When $\prescript{e \!}{}{K_{n\alpha i}^{m\beta j}}$ as given by (\ref{e30}) -- (\ref{e33}) is substituted into (\ref{e5}) for $K^{m\beta j}_{n\alpha i}$ one obtains the embedding-energy part of the dynamical matrix $\prescript{e}{}{\D}^{\beta j}_{\alpha i}(\K)$.  We have thus completed the determination of $\mathbb{D}$ in the most general case.

Nelson \textit{et al.} \cite{NelsonmPRB1989} have published expressions for ${K_{n\alpha i}^{m\beta j}}$ when $m\beta \ne n\alpha$; their expressions are consistent with the $m\beta \ne n\alpha$ terms in (\ref{e18}) and (\ref{e31}) -- (\ref{e33}).  Nelson \textit{et al.} do not directly consider the $n\alpha = m\beta$ FCs.  Their equations might therefore seem incomplete.  However, knowing the $m\beta \ne n\alpha$ FCs is sufficient, as Newton's third law allow one to readily find find ${K_{n\alpha i}^{n\alpha j}}$  via $K_{n\alpha i}^{n\alpha j} = - \sum_{m\beta (\ne n\alpha)} K_{n\alpha i}^{m\beta j}$.  This fact is readily apparent in (\ref{e18}), (\ref{e31}), and (\ref{e32}), where we see the $m\beta = n\alpha$ FC is indeed the negative of the sum over the $m\beta \ne n\alpha$ FCs.  Owing to simplifications that occur in the derivation of (\ref{e33}), the analogous relationship between the $m\beta = n\alpha$ and $m\beta \ne n\alpha$ FCs is not so readily apparent in this equation.

\subsection{Effective Pair Potentials}

The EAM energy expressed by (\ref{e1}) -- (\ref{e3}) has the interesting property that the division of $E$ into the components $E^p$ and $E^e$ is not unique.  Indeed, if we define the transformed pair-potentials
\be{\eq}
\label{e33a}
\bar{\phi}_{n \alpha}^{m \beta}  = \phi_{n \alpha}^{m \beta} + A_{m \beta} f_{ n\alpha}^{m \beta} + A_{n \alpha} f_{m \beta}^{n \alpha}
\en{\eq}
and embedding energies
\be{\eq}
\label{e33b}
\bar{F}_{n\alpha} = F_{n \alpha} - A_{n \alpha} \,\, \rho(\R_{n \alpha}),
\en{\eq}
then it is straightforward to show the total energy $E$ is unchanged.  Here $A_{n \alpha}$ is a constant associated with the atom designated by the subscript $n \alpha$.  Because $A_{n\alpha}$ can be different for each atom in the solid, the transformation defined by (\ref{e33a}) and (\ref{e33b}) can be viewed as a local gauge transformation.   

A particularly useful transformation occurs if we choose $A_{n \alpha} = D\!F_{n \alpha}$ where (as above) $D\!F_{n \alpha}$ is the derivative of $F_{n \alpha}$ with respect to its argument $\rho(\R_{n \alpha})$ evaluated at the equilibrium positions of the atoms in the material.  Then (\ref{e33a}) and (\ref{e33b}) become
\be{\eq}
\label{e33aa}
\bar{\phi}_{n \alpha}^{m \beta}  = \phi_{n \alpha}^{m \beta} + D \! F_{m \beta} f_{ n\alpha}^{m \beta} + D \! F_{n \alpha} f_{m \beta}^{n \alpha}
\en{\eq}
and
\be{\eq}
\label{e33bb}
\bar{F}_{n\alpha} = F_{n \alpha} - D \! F_{n \alpha} \,\, \rho(\R_{n \alpha}).
\en{\eq}
If we now calculate $D \! \bar{F}_{n \alpha}$ [the derivative of $\bar{F}_{n \alpha}$ with respect to the argument $\rho(\R_{n \alpha})$ evaluated at equilibrium], we straightforwardly obtain the simple result
\be{\eq}
\label{e33c}
D\!\bar{F}_{n\alpha} = 0.
\en{\eq}
Because the transformation expressed by (\ref{e33aa}) and (\ref{e33bb}) can be applied to any EAM model, all EAM models can be put on equal footing.  Indeed, embedded-atom-method models with the property $D \! \bar{F}_{n \alpha} = 0$ are known as normalized \cite{JOJMR1989}.  Owing to $\bar{\phi}_{n \alpha}^{m \beta}$ containing all pair-like interactions, the transformed potentials $\bar{\phi}_{n \alpha}^{m \beta}$ are often referred to as effective pair potentials.  In fact, a number of EAM models found in the literature impose $D\! {F}_{n\alpha} = 0$ from the outset \cite{JOJMR1989, WBJAP1995, Wilson2012, Chantas1996, Ercolessi1988, Johnson1990, Mishin2002}.   

The possibility of normalizing any EAM model is manifest in our above equations for $\prescript{p \!}{}{K_{n\alpha i}^{m\beta j}}$ and $\prescript{e \!}{}{K_{n\alpha i}^{m\beta j}}$.  Notice the sum of the embedding force-constant components $\prescript{e1 \!}{}{K_{n\alpha i}^{m\beta j}}$ and $\prescript{e2 \!}{}{K_{n\alpha i}^{m\beta j}}$ [see (\ref{e31}) and (\ref{e32})] is of the same form as the pair-potential constant $\prescript{p \!}{}{K_{n\alpha i}^{m\beta j}}$ [see (\ref{e18})].  Indeed, the sum of (\ref{e18}), (\ref{e31}), and (\ref{e32}) can be succinctly expressed in terms of the effective pair potentials $\bar{\phi}_{n \alpha}^{m \beta}$ as
\begin{align}
\label{e18b}
\prescript{ep \!}{}{K_{n\alpha i}^{m\beta j}} =  &\sum_{m'\beta'}  \Bigg[  \frac{D\bar{\phi}^{m\beta}_{m'\beta'}}{\rcur^{m\beta}_{m'\beta'}} \,  \Big( \delta_{ij} - \hrcur^{m\beta}_{m'\beta' i} \, \hrcur^{m\beta}_{m'\beta' j} \Big) +  D^2\bar{\phi}^{m\beta}_{m'\beta'}  \hrcur^{m\beta}_{m'\beta' i} \, \hrcur^{m\beta}_{m'\beta' j} \Bigg] \, \delta_{nm} \delta_{\alpha \beta} \nonumber \\ 
& - \Bigg[  \frac{D\bar{\phi}^{m\beta}_{n\alpha}}{\rcur^{m\beta}_{n\alpha}} \,  \Big( \delta_{ij} - \hrcur^{m\beta}_{n\alpha i} \, \hrcur^{m\beta}_{n\alpha j} \Big) +  D^2\bar{\phi}^{m\beta}_{n\alpha}  \hrcur^{m\beta}_{n\alpha i} \, \hrcur^{m\beta}_{n\alpha j} \Bigg] \, (1 - \delta_{nm} \delta_{\alpha \beta}), 
\end{align}
Insofar as $\prescript{e3 \!}{}{K_{n\alpha i}^{m\beta j}}$ [see (\ref{e33})] cannot be subsumed into a transformed pair-potential FC, it can be surmised that $\prescript{e3 \!}{}{K_{n\alpha i}^{m\beta j}}$ is the only part of  ${K_{n\alpha i}^{m\beta j}}$ uniquely attributable to many-body interactions.

\section{Application to Bulk bcc and fcc Materials}

Quite obviously, the general results for the FCs and resulting dynamical matrix are rather complicated.  Sometimes these equations must be applied in their full glory, as when using a slab in order to study vibrations near the surface of a material.

However, when calculating the bulk dynamics of a cubic lattice with one atom per unit cell -- as is the case of a bcc or fcc material -- a significant number of simplifications occur.  (i) Because there is only one atom per unit cell we may drop the indices ($\alpha$, $\beta$, etc.) that label the atoms within each unit cell.  (ii) Because all atoms are equivalent, the derivatives of $F$ are the same for each atom; we thus define the constants $F' = D\!F_m$ and $F'' = D^2\!F_m$.  (iii) Because all of the atoms are the same, we necessarily have $D\!f_n^m = D\!f_m^n$ and  $D^2\!f_n^m = D^2\!f_m^n$.  (iv)  Because the bcc and fcc lattices have inversion symmetry, the first three terms in (\ref{e33}) are each identically zero.  Taking these features into account, the equations for the force constants [(\ref{e18b}) and (\ref{e33})] respectively simplify to
\begin{align}
\label{e19}
\prescript{ep \!}{}{K_{n i}^{m j}} =  &\sum_{m'}  \bigg[  \frac{D\bar{\phi}^{m}_{m'}}{\rcur^{m}_{m'}}  \big( \delta_{ij} - \hrcur_{m' i}^{m} \, \hrcur_{m' j}^{m}  \big) +  D^2\bar{\phi}^{m}_{m'}  \hrcur^{m}_{m' i} \, \hrcur^{m}_{m' j} \bigg] \, \delta_{nm} \nonumber \\ 
& - \bigg[  \frac{D\bar{\phi}^{m}_{n}}{\rcur^{m}_{n}}  \big( \delta_{ij} - \hrcur_{n i}^{m} \, \hrcur_{n j}^{m}  \big) +  D^2\bar{\phi}^{m}_{n}  \hrcur^{m}_{n i} \, \hrcur^{m}_{n j} \bigg] \, (1 - \delta_{nm} ),
\end{align}
and
\begin{equation}
\label{e37}
\prescript{e3 \!}{}{K_{n i}^{m j}} = F'' \sum_{m'}  D\!f_{n}^{m'} \, D\!f^{m'}_{m} \, \hrcur_{n i}^{m'} \, \hrcur_{m j}^{m'} .
\end{equation}
The effective pair potential between any two atoms is now succinctly expressed as 
\be{\eq}
\label{e38}
\bar{\phi}_n^m(r) = \phi_n^m(r) + 2 F' f_n^m(r).
\en{\eq}

With these simplifications for the FCs, a fairly simple form for the dynamical matrix follows.  Inserting (\ref{e19}) for $\prescript{ep \!}{}{K_{n i}^{m j}}$ into (\ref{e5}) yields the effective pair-potential contribution to the dynamical matrix
\be{\eq} 
\label{e20}
\prescript{ep }{}\D_{i}^{j}(\K)  = \frac{1}{M} \sum_m \bigg[  \frac{D\bar{\phi}^{m}_{n}}{\rcur^{m}_{n}} \, \big( \delta_{ij} - \hrcur^{m}_{n i} \, \hrcur^{m}_{n j} \big) +  D^2\bar{\phi}^{m}_{n}  \hrcur^{m}_{n i} \, \hrcur^{m}_{n j} \bigg] \, \big[ 1 - \cos \! \big(\K 
\! \cdot \! (\R_{m} - \R_{n}) \big) \big],
\en{\eq}
where $M$ is the mass of each atom.  In writing this equation we have taken advantage of the relationships $\bar{\phi}^m_n = \bar{\phi}^n_m$, $\rcur^m_n = \rcur^n_m$, and $\hrcur^m_n = - \hrcur^n_m$.  We have also utilized the symmetry of the cubic lattice to explicitly eliminate the imaginary part of the dynamical matrix.  Because (\ref{e20}) only depends upon $\R_n$ via the displacement $\R_m - \R_n$, there is no dependence upon $n$;  this translational symmetry allows one to assume $\R_n$ is located at the origin.

We are left with finding the contribution of $\prescript{e3 \!}{}{K_{n i}^{m j}}$ to the dynamical matrix.  Inserting (\ref{e37}) into (\ref{e5}) readily gives us this remaining contribution,
\be{\eq}
\label{e39}
\prescript{e3}{}\D_{i}^{j}(\K) = \frac{F''}{M}  \sum_m \sum_{m'} D\!f_n^{m'} D\!f_m^{m'} \hrcur_{ni}^{m'} \hrcur_{mj}^{m'} \, e^{i \K  \cdot (\R_{m} - \R_n )}.
\en{\eq}
As it stands, this equation is a double sum on the two indices $m$ and $m'$.  We can simplify it to the product of two independent single sums (which is much faster to numerically compute) with a few manipulations.   We first  switch the order of the sums in (\ref{e39}) to yield
\be{\eq}
\label{e40}
\prescript{e3}{}\D_{i}^{j}(\K) = \frac{F''}{M} \sum_{m'} D\!f_n^{m'} \hrcur_{ni}^{m'} \, \sum_m D\!f_m^{m'} \hrcur_{mj}^{m'} \, e^{i \K \cdot (\R_{m} - \R_n )}.
\en{\eq}
We now define a new summation variable $m''$ for the interior sum via $m - m' = m'' - n$.  This gives us $\R_m = \R_{m''} + \R_{m'} - \R_n$, which allows us to rewrite (\ref{e40}) as
\be{\eq}
\label{e41}
\prescript{e3}{}\D_{i}^{j}(\K) = - \frac{F''}{M} \sum_{m'} D\!f_n^{m'} \hrcur_{ni}^{m'} e^{i \K \cdot (\R_{m'} - \R_n )} \, \sum_{m''} D\!f_n^{m''} \hrcur_{nj}^{m''} \, e^{i \K \cdot (\R_{m''} - \R_n )}.
\en{\eq}
Notice this equation is indeed the product of two independent sums.  The symmetry of the lattice allows further simplification, as only the terms containing the product $\sin(\K(\R_{m'} - \R_n)) \sin(\K(\R_{m''} - \R_n))$ yields a nonzero contribution.  We can thus write our final form for this part of the dynamical matrix as
\be{\eq}
\label{e42}
\prescript{e3}{}\D_{i}^{j}(\K) = \frac{F''}{M} \sum_{m'} D\!f_n^{m'} \hrcur_{ni}^{m'} \sin \! \big(\K \! \cdot \!  (\R_{m'} - \R_n ) \big) \, \sum_{m} D\!f_n^{m} \hrcur_{nj}^{m} \, \sin \! \big(\K \! \cdot \!  (\R_{m} - \R_n ) \big).
\en{\eq}
Together (\ref{e20}) and (\ref{e42}) represent the total dynamical matrix for a single-atom-basis material with cubic symmetry.

We note the result represented by (\ref{e20}) and (\ref{e42}) is consistent with that reported by Ningsheng \textit{et al.} \cite{NINGSHENGssc1989}, who pointed out the equations published by Daw and Hatcher \cite{DAWssc1985} were missing a factor of $2$ in the terms containing $F'$ [see (\ref{e38})].  The uncorrected result of Daw and Hatcher has apparently been used by Kaznac \textit{et al.} on two occasions \cite{KAZANCphysicab2005,KAZANCphysicab2006}.  We further note (\ref{e20}) and (\ref{e42}) are consistent with the expression for $\D$ given by Wang and Boercker \cite{WBJAP1995}.

\section{Relationship to BVK force constants}

In this section we consider Born-von-K\'{a}rm\'{a}n FCs in the context of the embedded-atom-method formalism.  We first discuss the relationship of BvK FCs to the EAM formalism.  We then utilize vibrational spectra from K, Fe, and Au to explore how the BvK - EAM relationship can be useful in evaluating the suitability of a given EAM model for modeling vibrational spectra.

\subsection{BvK force constants}

Experimental dispersion curves are commonly fit to extract what are known as Born-von-K\'{a}rm\'{a}n FCs, which we now briefly describe \cite{Shukla1966,Johnston1992}.  First, one assumes the equilibrium position of one particular atom is located at the origin.  Owing to the symmetry of the lattice, there is a minimal set of parameters that are required to describe all of the FCs between the atom at the origin and all of the atoms in a particular neighboring shell.  These parameters are often taken to be the BvK FCs.  For example, in a bcc lattice there are two independent BvK FCs ($\alpha^1_1$ and $\beta^1_1$) that can be used to describe the interactions between the atom at the origin and any atom in the first neighboring shell.  Generically, the BvK force-constant matrix $\prescript{bvk}{}\KK_k$ for the atom in the $k$th shell located at $\frac{1}{2} a_0 (h_x^k,h_y^k,h_z^k)$ where $h_x^k \ge h_y^k \ge h_z^k$ is assumed ($a_0$ is the lattice constant) is given by
\be{\eq}
\label{e49}
\prescript{bvk}{}\KK_k = \begin{pmatrix}
\alpha_1^k & \beta_3^k & \beta_2^k \\
\beta_3^k & \alpha_2^k & \beta_1^k \\
\beta_2^k & \beta_1^k & \alpha_3^k \\
\end{pmatrix} .
\en{\eq}
The force-constant matrix for other atoms in the $k$th shell are readily determined from this matrix via lattice symmetry \cite{White1958,Johnston1992}.  Specific BvK FC matrices for atoms in the first five shells of both bcc and fcc lattices are given in Table \ref{table1} \cite{Shukla1966,Johnston1992}.

\begin{table}[h] 
\caption{Definitions of Born-von-K\'{a}rm\'{a}n force-constant matrices $\prescript{bvk}{}\KK_k$ for specific atoms in the first five shells ($k = 1$ to 5) of bcc and fcc lattices.}  
\centering
\tabcolsep=0.12cm
\begin{tabular}{ccccc} 

\hline\hline
 Shell ($k$) & atom (bcc) &$\prescript{bvk}{}\KK_k$ (bcc) & atom (fcc) &$\prescript{bvk}{}\KK_i$ (fcc) \\

\hline
 
1 &$\frac{1}{2} a_0 (1,1,1)$ & $\begin{pmatrix}
                                          \alpha_1^1 & \beta_1^1 & \beta_1^1 \\
                                          \beta_1^1 & \alpha_1^1 & \beta_1^1 \\
                                          \beta_1^1 & \beta_1^1 & \alpha_1^1 \\
                                          \end{pmatrix}$ \vspace{1mm}         	&$\frac{1}{2} a_0 (1,1,0)$ 		& $\begin{pmatrix}
                                          														\alpha_1^1 & \beta_3^1 & 0 \\
                                          														\beta_3^1 & \alpha_1^1 & 0 \\
                                          														0 & 0 & \alpha_3^1 \\
                                          														\end{pmatrix}$ \vspace{1mm}         \\
 
2 &$\frac{1}{2} a_0 (2,0,0)$ & $\begin{pmatrix}
                                          \alpha_1^2 & 0 & 0 \\
                                          0 & \alpha_2^2 & 0 \\
                                          0 & 0 & \alpha_2^2 \\
                                          \end{pmatrix}$  \vspace{1mm}         &$\frac{1}{2} a_0 (2,0,0)$ 		& $\begin{pmatrix}
                                          														\alpha_1^2 & 0 & 0 \\
                                          														0 & \alpha_2^2 & 0 \\
                                          														0 & 0 & \alpha_2^2 \\
                                          														\end{pmatrix}$ \vspace{1mm}         \\

3 & $\frac{1}{2} a_0 (2,2,0)$ & $\begin{pmatrix}
                                          \alpha_1^3 & \beta_3^3 & 0 \\
                                          \beta_3^3 & \alpha_1^3 & 0 \\
                                          0 & 0 & \alpha_3^3 \\
                                          \end{pmatrix}$ \vspace{1mm}        &$\frac{1}{2} a_0 (2,1,1)$ 		& $\begin{pmatrix}
                                          														\alpha_1^3 & \beta_2^3 & \beta_2^3 \\
                                          														\beta_2^3 & \alpha_2^3 & \beta_1^3 \\
                                          														\beta_2^3 & \beta_1^3 & \alpha_2^3 \\
                                          														\end{pmatrix}$ \vspace{1mm}         \\

4 &$\frac{1}{2} a_0 (3,1,1)$ & $\begin{pmatrix}
                                          \alpha_1^4 & \beta_2^4 & \beta_2^4 \\
                                          \beta_2^4 & \alpha_2^4 & \beta_1^4 \\
                                          \beta_2^4 & \beta_1^4 & \alpha_2^4 \\
                                          \end{pmatrix}$ \vspace{1mm}        &$\frac{1}{2} a_0 (2,2,0)$ 		& $\begin{pmatrix}
                                          														\alpha_1^4 & \beta_3^4 & 0 \\
                                          														\beta_3^4 & \alpha_1^4 & 0 \\
                                          														0 & 0 & \alpha_3^4 \\
                                          														\end{pmatrix}$ \vspace{1mm}         \\

5 &$\frac{1}{2} a_0 (2,2,2)$ & $\begin{pmatrix}
                                          \alpha_1^5 & \beta_1^5 & \beta_1^5 \\
                                          \beta_1^5 & \alpha_1^5 & \beta_1^5 \\
                                          \beta_1^5 & \beta_1^5 & \alpha_1^5 \\
                                          \end{pmatrix}$ \vspace{1mm}        &$\frac{1}{2} a_0 (3,1,0)$ 		& $\begin{pmatrix}
                                          														\alpha_1^5 & \beta_3^5 & 0 \\
                                          														\beta_3^5 & \alpha_2^5 & 0 \\
                                          														0 & 0 & \alpha_3^5 \\
                                          														\end{pmatrix}$ \vspace{1mm}         \\

\hline\hline

\end{tabular}
\label{table1}
\end{table}

\begin{table}[b] 
\caption{First-shell through fifth-shell bcc-lattice Born-von-K\'{a}rm\'{a}n force-constants for a normalized EAM model with interactions [$\phib(r)$ and $f(r)$] that extend to fifth-shell neighbors.  The numbers in the column labeled ``Index'' are used in part (e) of Figs.~1 and 2 to denote the particular BvK FC.}  
\centering
\tabcolsep=0.11cm
\begin{tabular}{rrll} 

\hline\hline

Shell & Index & Pair-potential FCs  & Embedding-energy FCs  \\

\hline
 
\begin{tabular}{l}
1 \\	
$\,$ \\
$\,$ 
\end{tabular} 

& \begin{tabular}{l}
1 \\	
$\,$ \\
2
\end{tabular}

	&\begin{tabular}{l}
      	$\alpha_{1p}^1 = \frac{2}{3} \phib'_1/r_1 + \frac{1}{3} \phib_1''$ \\
  	$\,$ \\
	$\beta_{1p}^1 = -\frac{1}{3} \phib'_1/r_1 + \frac{1}{3} \phib_1''$ 
	
	\end{tabular}  										&\begin{tabular}{l}
                             										$\alpha_{1e}^1 = - F'' \big[\frac{2}{3} f'_1 \big(\sqrt{3} f'_2 + \sqrt{6} f'_3  + f'_5 \big)$ \\ 
													$\quad \quad \quad \!+ \frac{6}{\sqrt{11}} f'_4 \big( f'_2 + \frac{4 \sqrt{2}}{3} f'_3 +\frac{5 }{3\sqrt{3}} f'_5 \big)\big]$ \\
  													$\beta_{1e}^1 =  F'' \big[ \frac{2}{3} f'_1 \big(\sqrt{3} f'_2 - f'_5 \big) - \frac{2}{\sqrt{11}} f'_4 \big( f'_2 - 2 \sqrt{2} f'_3 -  \sqrt{3} f'_5 \big) \big]$
													\end{tabular}\\
\hline 

\begin{tabular}{l}
2 \\	
$\,$ 
\end{tabular}

& \begin{tabular}{l}
3 \\	
4
\end{tabular}

 	&  \begin{tabular}{l}
      	$\alpha_{1p}^2 =  \phib_2''$\\
  	$\alpha_{2p}^2 =  \phib'_2/r_2 $
	\end{tabular}  										& 	\begin{tabular}{l}
                             										$\alpha_{1e}^2 = F'' \big[\frac{4}{3} (f'_1)^2 - \frac{24 }{\sqrt{33}}  f'_1 f'_4 + \frac{8}{11} (f'_4)^2 \big]$\\
  													$\beta_{2e}^2 =   - F'' \big[ \frac{4}{3} (f'_1)^2 + \frac{8}{\sqrt{33}}  f'_1 f'_4  + 2 \sqrt{2} f'_3 \big( f'_2 + \frac{2}{\sqrt{3}} f'_5 \big) + \frac{40}{11} (f'_4)^2 \big]$
													\end{tabular}\\
																								
\hline 
																																											
\begin{tabular}{l}
3 \\	
$\,$ \\
$\,$
\end{tabular}

& \begin{tabular}{l}
5 \\	
6 \\
7
\end{tabular}

 	&  \begin{tabular}{l}
      	$\alpha_{1p}^3 = \frac{1}{2} \phib'_3/r_3 + \frac{1}{2} \phib_3''$\\
  	$\alpha_{3p}^3 =  \phib'_3/r_3 $\\
	$\beta_{3p}^3 = - \frac{1}{2} \phib'_3/r_3 + \frac{1}{2} \phib_3''$
	\end{tabular}  										& 	\begin{tabular}{l}
                             										$\alpha_{1e}^3 = F'' \big[ \frac{2}{3} (f'_1)^2 - \frac{8}{\sqrt{33}}  f'_1 f'_4 - \frac{10}{11} (f'_4)^2 \big]$\\
  													$\alpha_{3e}^3 =  -F'' \big[ \frac{2}{3} (f'_1)^2 + \frac{8}{\sqrt{33}}  f'_1 f'_4  + \frac{4}{\sqrt{3}} f'_2 f'_5 + 2 (f'_3)^2 + 2 (f'_4)^2 \big]$\\
													$\beta_{3e}^3 = F'' \big[ \frac{2}{3} (f'_1)^2 + \frac{8}{\sqrt{33}}  f'_1 f'_4  + (f'_2)^2 + (f'_3)^2 + 2 (f'_4)^2 \big]$
													\end{tabular}\\

\hline

\begin{tabular}{l}
4 \\	
$\,$ \\
$\,$ \\
$\,$
\end{tabular}

& \begin{tabular}{l}
$\,\,\,$8 \\	
$\,\,\,$9 \\
10 \\
11
\end{tabular}

       &  \begin{tabular}{l}
         $\alpha_{1p}^4 = \frac{2}{11} \phib'_4/r_4 + \frac{9}{11} \phib_4''$\\
  	$\alpha_{2p}^4 =  \frac{10}{11} \phib'_4/r_4 + \frac{1}{11} \phib_4''$\\
	$\beta_{1p}^4 = - \frac{1}{11} \phib'_4/r_4 + \frac{1}{11} \phib_4''$ \\
	$\beta_{2p}^4 = - \frac{3}{11} \phib'_4/r_4 + \frac{3}{11} \phib_4''$
	\end{tabular}  										& 	\begin{tabular}{l}
                             										$\alpha_{1e}^4 = F'' \big[ \frac{2}{3} f'_1 \big(\sqrt{3} f'_2 + \sqrt{6} f'_3 + f'_5  \big)  + \frac{2}{\sqrt{11}} f'_4 \big( \sqrt{2} f'_3 + \frac{2}{\sqrt{3}} f'_5 \big) \big]$\\
  													$\alpha_{2e}^4 =  - F'' \big[\frac{1}{3} f'_1 \big( \sqrt{6} f'_3  + 2 f'_5 \big) + \frac{2}{\sqrt{11}} f'_4 \big( f'_2 + 2 \sqrt{2} f'_3 +\frac{4}{\sqrt{3}} f'_5 \big)\big] $\\
													$\beta_{1e}^4 = F'' \big[\frac{1}{3} f'_1 \big( \sqrt{6} f'_3  - 2 f'_5 \big) + \frac{2}{\sqrt{11}} f'_4 \big( f'_2 - \sqrt{2} f'_3 +\frac{4}{\sqrt{3}} f'_5 \big)\big]$\\
													$\beta_{2e}^4 = F'' \big[\frac{1}{3} f'_1 \big( \sqrt{3} f'_2  + \frac{\sqrt{3}}{\sqrt{2}} f'_3 \big) + \frac{3}{\sqrt{11}} f'_4 \big( f'_2 +  \sqrt{2} f'_3 +\frac{2}{3 \sqrt{3}} f'_5 \big)\big] $
													\end{tabular}\\

\hline

\begin{tabular}{l}
5 \\	
$\,$
\end{tabular}

& \begin{tabular}{l}
12 \\	
13
\end{tabular}

     	&  \begin{tabular}{l}
      	$\alpha_{1p}^5 = \frac{2}{3} \phib'_5/r_5 + \frac{1}{3} \phib_5''$\\
  	$\beta_{1p}^5 = -\frac{1}{3} \phib'_5/r_5 + \frac{1}{3} \phib_5''$
	\end{tabular}  										& 	\begin{tabular}{l}
                             										$\alpha_{1e}^5 = F'' \big[ \frac{1}{3} (f'_1)^2 - \frac{2}{\sqrt{33}}  f'_1 f'_4 - \frac{10}{11} (f'_4)^2 \big]$\\
  													$\beta_{1e}^5 =  F'' \big[ \frac{1}{3} (f'_1)^2 + \frac{6}{\sqrt{33}}  f'_1 f'_4 + \sqrt{2} f'_2 f'_3 + \frac{14}{11} (f'_4)^2 \big] $
													\end{tabular}\\

\hline\hline

\end{tabular}


\label{table2}
\end{table}

\begin{table}[h] 
\caption{First-shell through fifth-shell fcc-lattice Born-von-K\'{a}rm\'{a}n force-constants for a normalized EAM model with interactions [$\phib(r)$ and $f(r)$] that extend to fifth-shell neighbors. The numbers in the column labeled ``Index'' are used in part (e) of Fig.~3 to denote the particular BvK FC. }  
\centering
\tabcolsep=0.11cm

\begin{tabular}{rrll} 

\hline\hline

Shell & Index & Pair-potential FCs  & Embedding-energy FCs  \\

\hline
 
\begin{tabular}{l}
1 \\	
$\,$ \\
$\,$ \\
$\,$ \\
$\,$ \\
$\,$ 
\end{tabular} 

& \begin{tabular}{l}
1 \\
$\,$ \\	
2 \\
$\,$ \\
3 \\
$\,$ \\
\end{tabular}

	&\begin{tabular}{l}
      	$\alpha_{1p}^1 = \frac{1}{2} \phib'_1/r_1 + \frac{1}{2} \phib_1''$ \\
	$\,$ \\
  	$\alpha_{3p}^1 = \phib'_1/r_1 $ \\
	$\,$ \\
	$\beta_{3p}^1 = - \frac{1}{2} \phib'_1/r_1 + \frac{1}{2} \phib_1''$ \\
	$\,$ 
	\end{tabular}  										&\begin{tabular}{l}
                             										$\alpha_{1e}^1 = - F'' \big[ f'_1 \big(\sqrt{2} f'_2 + \frac{4}{\sqrt{3}} f'_3  + f'_4 \big) + \frac{4}{3} f'_3 \big( f'_3 + \frac{\sqrt{3}}{2} f'_4 + \frac{3 \sqrt{3}}{\sqrt{5}} f'_5 \big)$ \\ 
													$\quad \quad \quad \!+ \frac{3}{\sqrt{5}} f'_5 \big(\sqrt{2} f'_2 + \frac{4}{3} f'_4 \big) \big]$ \\
  													$\alpha_{3e}^1 = - F'' \big[ f'_1 \big(2 f'_1 + \frac{4}{\sqrt{3}} f'_3 \big) + \frac{4}{3} f'_3 \big(\sqrt{6} f'_2 + \frac{1}{2} f'_3 + 2 \sqrt{3} f'_4 + \frac{\sqrt{3}}{\sqrt{5}} f'_5 \big)$ \\ 
													$\quad \quad \quad \! + \frac{18}{5} (f'_5)^2 \big]$ \\
													$\beta_{3e}^1 = F'' \big[ f'_1 \big( f'_1 + \sqrt{2} f'_2 - \frac{2}{\sqrt{3}} f'_3  - f'_4 \big) + \frac{2}{3} f'_3 \big( \frac{5}{2} f'_3 + \sqrt{3} f'_4 + \frac{3 \sqrt{3}}{\sqrt{5}} f'_5 \big)$ \\ 
													$\quad \quad \quad \! - \frac{1}{\sqrt{5}} f'_5 \big( \sqrt{2} f'_2 - 4 f'_4 - \frac{1}{\sqrt{5}} \ f'_5  \big) \big]$ 
													\end{tabular}\\
\hline 

\begin{tabular}{l}
2 \\	
$\,$ 
\end{tabular}

& \begin{tabular}{l}
4 \\	
5
\end{tabular}

 	&  \begin{tabular}{l}
      	$\alpha_{1p}^2 =  \phib_2''$\\
  	$\alpha_{2p}^2 =  \phib'_2/r_2 $
	\end{tabular}  										& 	\begin{tabular}{l}
                             										$\alpha_{1e}^2 = F'' \big[2 f'_1 \big( f'_1 -\frac{6}{\sqrt{5}} f'_5 \big) + \frac{4}{3} (f'_3)^2 + \frac{2}{5} (f'_5)^2 \big]$\\
  													$\alpha_{2e}^2 =   - F'' \big[f'_1 \big( f'_1 + \frac{4}{\sqrt{3}} f'_3 + \frac{2}{\sqrt{5}} f'_5 \big) + 2 \sqrt{2} f'_2 f'_4  + \frac{10}{3} (f'_3)^2 + \frac{9}{5} (f'_5)^2 \big]$
													\end{tabular}\\
																								
\hline 
																																											
\begin{tabular}{l}
3 \\	
$\,$ \\
$\,$ \\
$\,$ \\
$\,$ \\
$\,$ \\
$\,$
\end{tabular}

& \begin{tabular}{l}
6 \\	
7 \\
$\,$ \\
8 \\
$\,$ \\
9\\
$\,$ 
\end{tabular}

 	&  \begin{tabular}{l}
      	$\alpha_{1p}^3 = \frac{1}{3} \phib'_3/r_3 + \frac{2}{3} \phib_3'' $ \\
  	$\alpha_{2p}^3 =  \frac{5}{6} \phib'_3/r_3 + \frac{1}{6} \phib_3''$ \\
	$\,$ \\
	$\beta_{1p}^3 = - \frac{1}{6} \phib'_3/r_3 + \frac{1}{6} \phib_3''$ \\
	$\,$ \\
	$\beta_{2p}^3 = - \frac{1}{3} \phib'_3/r_3 + \frac{1}{3} \phib_3''$ \\
	$\,$ \\
	\end{tabular}  										& 	\begin{tabular}{l}
                             										$\alpha_{1e}^3 =  F'' \big[ f'_1 \big( f'_1  + \frac{2}{\sqrt{3}} f'_3  - \frac{6}{\sqrt{5}} f'_5 \big) + \frac{2}{3} f'_3 \big(\frac{1}{2} f'_3 - \frac{2 \sqrt{3}}{\sqrt{5}} f'_5 \big) \big]$ \\ 
													$\alpha_{2e}^3 = - F'' \big[ f'_1 \big(\frac{2}{\sqrt{3}} f'_3 + f'_4 \big) + \frac{2}{3} f'_3 \big( \frac{\sqrt{3}}{\sqrt{2}} f'_2 + f'_3 + \frac{\sqrt{3}}{2} f'_4 + \frac{4 \sqrt{3}}{\sqrt{5}} f'_5 \big)$ \\ 
													$\quad \quad \quad \! +  \frac{3}{\sqrt{5}} f'_4 f'_5 \big]$ \\
  													$\beta_{1e}^3 = F'' \big[ f'_1 \big(\frac{1}{2} f'_1 - \frac{1}{\sqrt{3}} f'_3  + f'_4 + \frac{1}{\sqrt{5}} f'_5\big)  $ \\ 
													$\quad \quad \quad \! + \frac{1}{3} f'_3 \big( \sqrt{6} f'_2 + \frac{5}{2} f'_3 - \sqrt{3} f'_4 + \frac{2 \sqrt{3}}{\sqrt{5}} f'_5 \big) - \frac{1}{\sqrt{5}} f'_4 f'_5 \big]$ \\
													$\beta_{2e}^3 = F'' \big[ f'_1 \big(\frac{1}{2} f'_1 + \frac{1}{\sqrt{2}} f'_2 + \frac{1}{\sqrt{3}} f'_3  + \frac{1}{\sqrt{5}} f'_5 \big)  $ \\ 
													$\quad \quad \quad \! + \frac{1}{3} f'_3 \big( \sqrt{6} f'_2 + \frac{1}{2} f'_3 + \sqrt{3} f'_4 + \frac{6 \sqrt{3}}{\sqrt{5}} f'_5 \big) + \frac{2}{\sqrt{5}} f'_4 f'_5 \big]$ \\ 
													\end{tabular}\\

\hline

\begin{tabular}{l}
4 \\	
$\,$ \\
$\,$ \\
$\,$
\end{tabular}

& \begin{tabular}{l}
10 \\	
11 \\
12 \\
$\,$
\end{tabular}

       &  \begin{tabular}{l}
         $\alpha_{1p}^4 = \frac{1}{2} \phib'_4/r_4 + \frac{1}{2} \phib_4''$\\
  	$\alpha_{3p}^4 =  \phib'_4/r_4 $\\
	$\beta_{3p}^4 = - \frac{1}{2} \phib'_4/r_4 + \frac{1}{2} \phib_4''$ \\
	$\,$ \\
	\end{tabular}  										& 	\begin{tabular}{l}
                             										$\alpha_{1e}^4 = F'' \big[ f'_1 \big( \frac{1}{2} f'_1 + \frac{2}{\sqrt{3}} f'_3 - \frac{2}{\sqrt{5}} f'_5  \big)  + \frac{2}{3} f'_3 \big(\frac{1}{2} f'_3 - \frac{3 \sqrt{3}}{\sqrt{5}} f'_5 \big) - \frac{3}{5} (f'_5)^2 \big]$\\
  													$\alpha_{3e}^4 =  - F'' \big[ \frac{4}{3} f'_3 \big(\sqrt{3} f'_1 + f'_3 + \frac{\sqrt{3}}{\sqrt{5}} f'_5 \big) + 2 (f'_4)^2   \big] $\\
													$\beta_{3e}^4 = F'' \big[ f'_1 \big(\frac{1}{2} f'_1  + \frac{4}{\sqrt{3}} f'_3  + \frac{2}{\sqrt{5}} f'_5 \big) + (f'_2)^2 $ \\ 
													$\quad \quad \quad \! + \frac{4}{3} f'_3 \big( \frac{1}{4} f'_3 +  \frac{3 \sqrt{3}}{\sqrt{5}} f'_5 \big) + (f'_4)^2 + (f'_5)^2 \big]$
													\end{tabular}\\

\hline

\begin{tabular}{l}
5 \\	
$\,$ \\
$\,$ \\
$\,$ \\
$\,$ 
\end{tabular}

& \begin{tabular}{l}
13 \\	
14 \\
15 \\
16 \\
$\,$ 
\end{tabular}

     	&  \begin{tabular}{l}
      	$\alpha_{1p}^5 = \frac{1}{10} \phib'_5/r_5 + \frac{9}{10} \phib_5''$\\
  	$\alpha_{2p}^5 = \frac{9}{10} \phib'_5/r_5 + \frac{1}{10} \phib_5''$ \\
	$\alpha_{3p}^5 = \phib'_5/r_5 $ \\
	$\beta_{3p}^5 = -\frac{3}{10} \phib'_5/r_5 + \frac{3}{10} \phib_5''$ \\
	$\,$ 
	\end{tabular}  										& 	\begin{tabular}{l}
                             										$\alpha_{1e}^5 = F'' \big[ f'_1 \big( \sqrt{2} f'_2 +\frac{4}{\sqrt{3}} f'_3 +f'_4  \big)  + \frac{4}{3} f'_3 \big( f'_3 + \frac{\sqrt{3}}{2} f'_4 \big) + \frac{1}{\sqrt{5}} f'_4 f'_5\big]$ \\
  													$\alpha_{2e}^5 =  -F'' \big[ f'_1 f'_4 + \frac{4}{3} (f'_3)^2 + \frac{1}{\sqrt{5}} f'_5 \big( \sqrt{2} f'_2 + 3 f'_4  \big)  \big] $ \\
													$\alpha_{3e}^5 =  -F'' \big[  2 f'_1 \big( \frac{1}{\sqrt{3}} f'_3 + \frac{1}{\sqrt{5}} f'_5  \big) + \frac{2}{3} f'_3 \big(  f'_3 + 2 \sqrt{3} f'_4 \big)  \big] $ \\
													$\beta_{3e}^5 =  F'' \big[ f'_1 \big( \frac{1}{\sqrt{2}} f'_2 + \frac{1}{\sqrt{3}} f'_3 + \frac{3}{\sqrt{5}} f'_5  \big) + \frac{1}{3} f'_3 \big ( 3 f'_3 + \sqrt{3} f'_4  \big) $ \\
													$ \quad \quad \quad \! + \frac{3}{\sqrt{5}} f'_5 \big( \frac{1}{\sqrt{2}} f'_2+ \frac{1}{3} f'_4  \big) \big] $ 
													\end{tabular}\\

\hline\hline

\end{tabular}
\label{table3}
\end{table}

So how are the BvK FCs related to the EAM FCs derived above?  Referring to Table \ref{table1}, we see (for example) that the $xy$ component of $\prescript{bvk}{}\KK_3$ (bcc) is designated $\beta_3^3$.  The BvK FCs are defined such that $\beta_3^3$ is the force on the atom at the origin in the $x$ direction when the atom located at $\frac{1}{2}a_0 (2,2,0)$ is displaced in the $y$ direction a unit distance.  Given (\ref{e6a}), this means the BvK FCs are the negative of our EAM FCs when the EAM force constants are applied to the appropriate pair of atoms.  For a bcc or fcc lattice we can thus directly use (\ref{e19}) and (\ref{e37}) to evaluate the BvK FCs in terms of the EAM model.  We note the pair-potential contribution comes from the second term in (\ref{e19}) as only this term is nonzero when $m \ne n$.

We have evaluated the BvK FCs for both lattice types assuming a normalized EAM model with functions $\bar{\phi}(r)$ and $f(r)$ that are nonzero out to the fifth shell of neighbors.  The results are displayed in Tables \ref{table2} and \ref{table3}.  In the service of clarity we have broken up each FC into pair-potential and embedding contributions: for example, $\alpha_1^1 = \alpha_{1p}^1  + \alpha_{1e}^1$.  As is evident in the tables the key quantities for each shell are the first and second derivatives $\phi'_k$ and  $\phi''_k$, respectively, of $\bar{\phi}(r)$ and the first derivative $f'_k$ of $f(r)$, where all derivatives are evaluated at the shell distances $r_k$.  Our equations for the pair-potential contribution to the bcc FCs agree with previously published expressions\cite{Cochran1963}.  As is evident in (\ref{e37}), the embedding part of the interaction between the $n$th and $m$th atoms is mediated by other nearby atoms in the lattice, making the effective range of the embedding interaction twice the distance of the range of $f(r)$.  This feature is manifest in the FCs in Tables \ref{table2} and \ref{table3}.  Notice, for example, in Table \ref{table2} terms with $(f'_1)^2$ appear in the FCs for the fifth bcc shell.


\subsection{Vibrational Spectra of K, Fe, and Au}

We now compare experimental and EAM-model-calculated vibrational spectra.  Specifically, we look at vibrations in K, Fe, and Au with an eye towards assessing which force constants are most important for an EAM model to accurately predict.  We chose to examine these three metals because (i) they have all been extensively modeled using the EAM formalism, (ii) simple, transition, and noble metals are each represented, (iii) both bcc and fcc lattice types are included, and (iv) high quality experimental dispersion curves  -- with concurrent BvK analysis -- have been published for all three metals.

\begin{figure}[b]
\label{figure1}\includegraphics[scale=0.72]{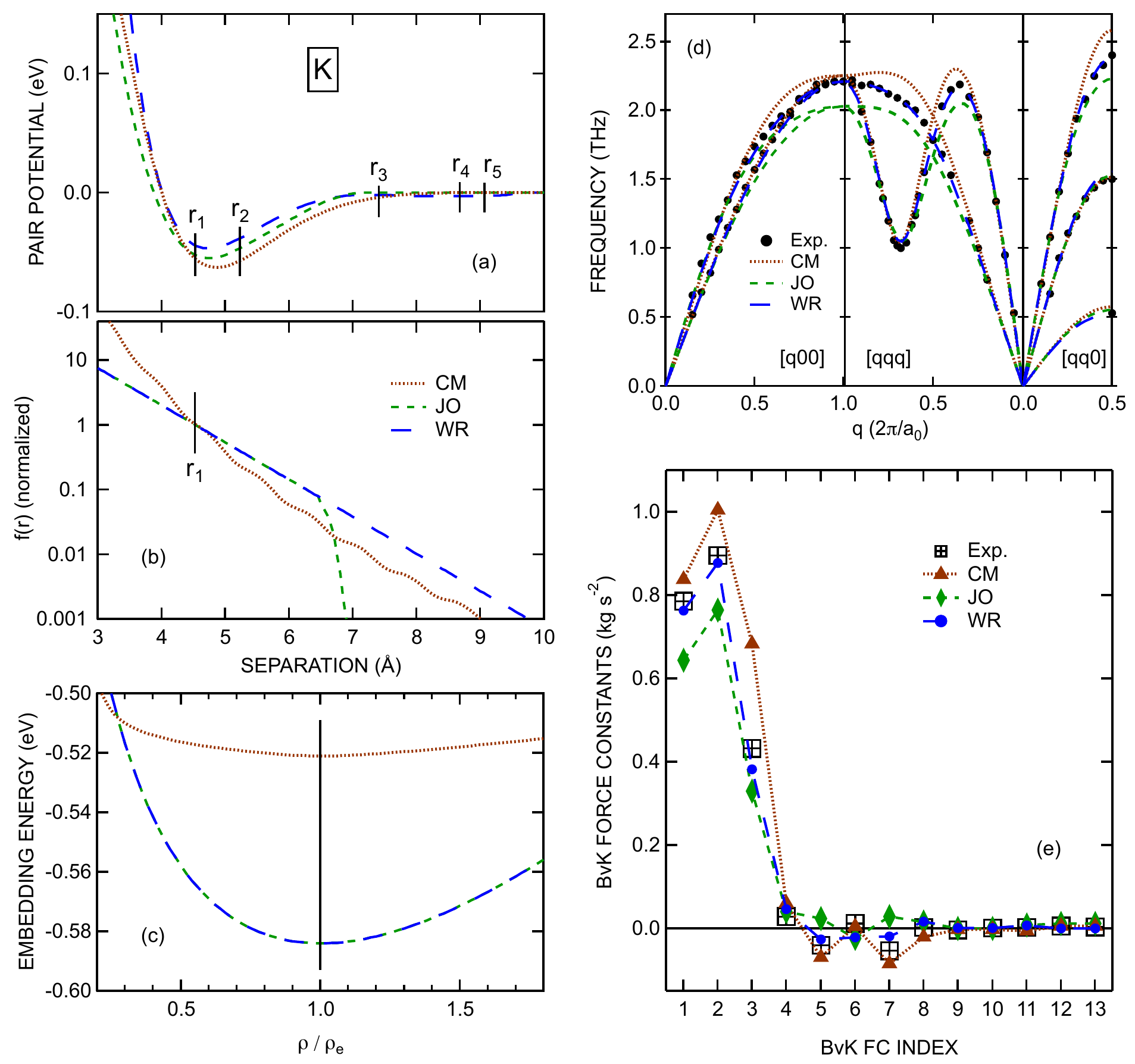}
\caption{Potassium EAM models, dispersion curves, and BvK FCs.  The effective potential energy $\bar{\phi}(r)$, atomic electron density $f(r)$, and normalized embedding energy $\bar{F}(\rho)$ are displayed in (a), (b), and (c), respectively; the dotted, short-dashed, and long-dashed curves corresponds to the EAM models of CM \cite{CMPRB1998}, JO \cite{JOJMR1989}, and WR \cite{Wilson2012}, respectively.  In (d) EAM-model derived dispersion curves are compared with the experimental data of Cowley \textit{et al.} \cite{CowleyPR1966}, while in (e) EAM-model calculated BvK force constants are compared with those extracted by Cowley \textit{et al.} from the experimental dispersion-curve data displayed in (d). }
\end{figure}

We start by looking at the EAM models for K of Chantasiriwan and Milstein (CM) \cite{CMPRB1998}, Johnson and Oh (JO) \cite{JOJMR1989}, and Wilson and Riffe (WR) \cite{Wilson2012}, which were previously compared in an EAM investigation of all five alkali metals \cite{Wilson2012}.  Parts (a), (b) and (c) of Fig.~1 illustrate the three defining functions [$\bar{\phi}(r)$, $f(r)$, and $\bar{F}(\rho)$] of each model.  It so happens each of these models is explicitly normalized; hence $\bar{\phi}(r) = \phi(r)$ and $\bar{F}(\rho) = F(\rho)$.  As shown in (a) the pair potentials of the three models are qualitatively similar, with an overall minimum between the first-neighbor distance $r_1$ and second-neighbor distance $r_2$.  The direct interactions in the CM and WR models extend out to the fifth shell of neighbors.  The JO pair potential and atomic charge density both go to zero for $r$ somewhere between the second and third neighbor distances, and so the JO model only includes direct interactions out to the second shell.

Experimental dispersion curves calculated using these three models are compared with the experimental data (solid circles) of Cowley \textit{et al.} \cite{CowleyPR1966} in part (d) of Fig.~1.  All three models do quite well near the zone center ($q=0$); this feature can be attributed to to the use of elastic constants in setting the parameters of each model.  However, away from the zone center the three models become distinguished, with the WR model providing a uniformly accurate accounting of the dispersion that the other two models lack.

Perhaps it is no surprise, then, that the BvK force constants calculated from the WR model best match those directly derived from the experimental data, as is evident in part (e) of Fig.~1.  Interestingly, the magnitudes of the first three BvK FCs ($\alpha_1^1$, $\beta_1^1$, and $\alpha_1^2$) are significantly larger than the remaining FCs.  Indeed, the FCs with FC Index $> 3$ have magnitudes that are less than $10 \%$ of $\alpha^2_1$, the smallest of the first three FCs.  These observations suggests an EAM model that accurately predicts the first three FCs while minimizing the absolute values of the remaining FCs might do very well at predicting K vibrational spectra.

We now move on to Fe.  In a study that focused on the ability of EAM models to predict surface relaxation, Haftel \textit{et al.} \cite{Haftel1990} introduced six different EAM models for Fe. \footnote{In implementing the models of Haftel \textit{et al.} we found several typos: (i) For model H6 $r_{cl}$ should be 3.5 rather than 3.9, (ii) for model H3 $v_4$ should be $-91621.76101$ rather than $-92621.76101$, and (iii) for model H6 $v_8$ should be $-102985.92764$ rather than $-102985.95764$.}  Following their numbering scheme, we present the three defining functions for all six models in (a), (b), and (c) of Fig.~2.   As can be see in (a) the effective pair potentials vary quite dramatically from one model to the next.  Models H3 and H5 include direct interactions out to the second shell of neighbors; the other four models also include the third shell.  The relative complexity of these pair potentials -- compared to those for K -- is likely attributable to the fact that Fe is a transition metal as opposed to a simple metal.

\begin{figure}[b]
\label{fig2}\includegraphics[scale=0.72]{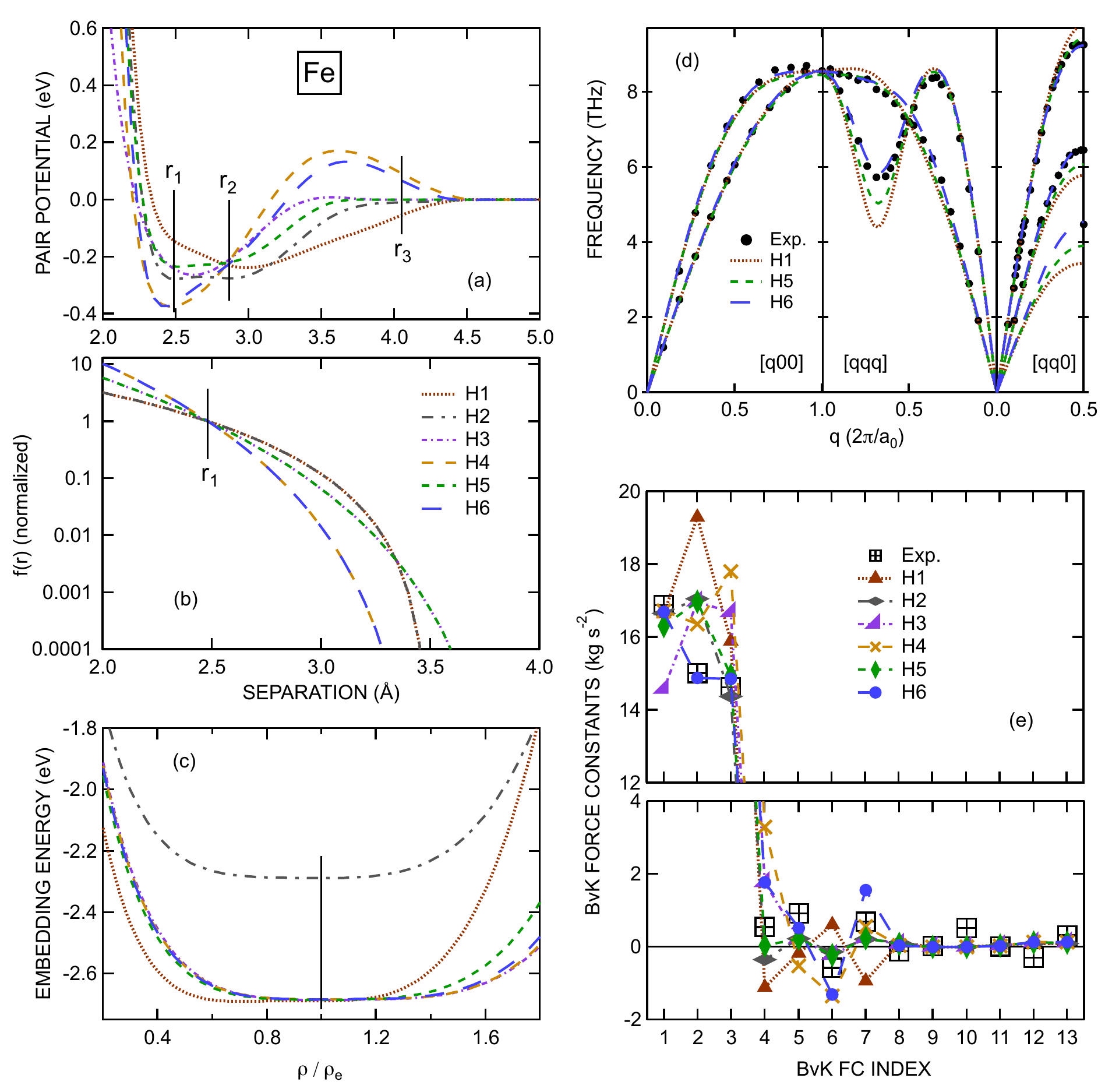}
\caption{Iron EAM models, dispersion curves, and BvK FCs.  The effective potential energy $\bar{\phi}(r)$, atomic electron density $f(r)$, and normalized embedding energy $\bar{F}(\rho)$ are displayed in (a), (b), and (c), respectively; these curves corresponds to the six EAM models presented by Haftel \textit{et al.} \cite{Haftel1990}.  In (d) EAM-model derived dispersion curves (from three Haftel \textit{et al.}  models) are compared with the experimental data of Minkiewicz \textit{et al.} \cite{Minkiewicz1967}, while in (e) EAM-model calculated BvK force constants are compared with those extracted by Minkiewicz \textit{et al.} from the experimental dispersion-curve data displayed in (d). }
\end{figure}

Dispersion curves calculated using three of these models (H1, H5, and H6) are compared against the experimental data of Minkiewicz \textit{et al.} \cite{Minkiewicz1967} in part (d) of Fig.~2.  The dispersion curves obtained from the other models of Haftel \textit{et al.} are similar to those shown here, and so have been omitted for clarity. As with K, all models do a good job near the zone center, again owing to use of elastic constants as input parameters in each model.  Overall, model H6 is the most accurate reproducing the experimental dispersion curves.

Our observations regarding the BvK FCs are largely the same as for K.  First, the first three FCs are again much larger than any of the remaining FCs, although the dominance is not quite as pronounced in the present case.  Second, these first three FCs are most accurately predicted by the model -- model H6 in this case -- that most accurately predicts the experimental dispersion curves.

It is worth closely comparing the FC results for models H5 and H6.  As is evident in Fig.~2(e), model H6 predicts the first three FCs quite well, but does less well with the next four constants ($\alpha^2_2$ and the three third-shell FCs).  In contrast, model H5 is much better overall at predicting the latter four FCs, but it does miss the mark as far as the second FC goes.  As Fig.~2(d) illustrates, model H5 does a much poorer job than model H6 with the dispersion curves.  These results emphasize the importance of accurately predicting the first three BvK FCs.  Furthermore, the results suggest that one might be able to find second-neighbor models for both K and Fe that can accurately describe vibrations in these two materials.

Lastly, we consider vibrations in Au.  In searching the literature we found six EAM models for Au \cite{Ercolessi1988, Mei1991, Cai1996, Pohlong1998, CMPRB1998, Sheng2011}.  In parts (a), (b), and (c) of Fig.~3 we plot the defining functions for the three EAM models that do the best job of reproducing the experimental dispersion curves of Lynn \textit{et al} \cite{Lynn1973}.  The models are from CM \cite{CMPRB1998}, Pohlong and Ram (PR) \cite{Pohlong1998}, and Sheng \textit{et al.} (SH) \cite{Sheng2011}.  The models of PR and CM include direct interactions out to the third shell of neighbors, while those of SH model extend to the fourth shell. 

\begin{figure}[t]
\label{fig3}\includegraphics[scale=0.72]{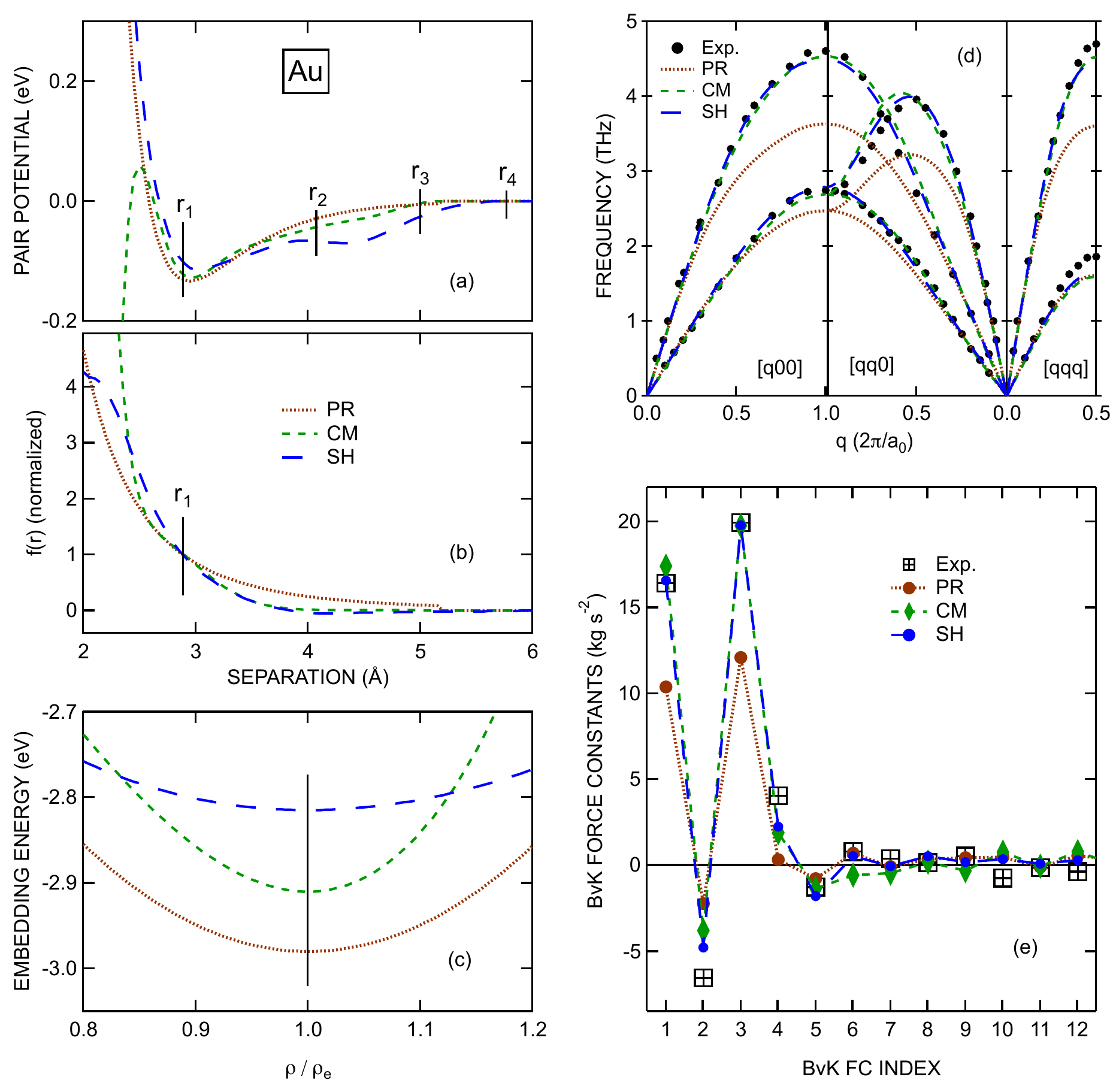}
\caption{Gold EAM models, dispersion curves, and BvK FCs.  The effective potential energy $\bar{\phi}(r)$, atomic electron density $f(r)$, and normalized embedding energy $\bar{F}(\rho)$ are displayed in (a), (b), and (c), respectively; the dotted, short-dashed, and long-dashed curves corresponds to the EAM models of PR \cite{Pohlong1998}, CM \cite{CMPRB1998}, and SH \cite{Sheng2011}, respectively.  In (d) EAM-model derived dispersion curves are compared with the experimental data of Lynn \textit{et al.} \cite{Lynn1973}, while in (e) EAM-model calculated BvK force constants are compared with those extracted by Lynn \textit{et al.} from the experimental dispersion-curve data displayed in (d). }
\end{figure}  

The dispersion-curve -- BvK-FC correlations for Au are not unlike those for K and Fe.  As Fig.~3(d) shows, the CM and SH models do almost equally well as matching the experimental dispersion curves of Lynn \textit{et al.}, and both are superior to the PR model.\footnote{Surprisingly, the dispersion curves we have calculated using the Sheng \textit{et al.} model match the experimental data significantly better than those presented in Ref.~[\onlinecite{Sheng2011}].  We discovered that we can reproduce the theoretical curves in Ref.~[\onlinecite{Sheng2011}] by neglecting the (normalized) embedding-energy contribution to the potential energy.}  Not surprisingly, the CM and SH models predict first-shell and second-shell FCs that are closest to those derived directly from the experimental dispersion curves, as is observed in part (e) of Fig.~3.  The relative magnitudes of the BvK FCs  for Au suggests a second-neighbor model might suffice to describe the interaction in this metal.

It is instructive to consider the relative contributions of the (effective) pair potential and embedding energy to the BvK FCs in these three metals.  For all K and Fe EAM models considered here the dominant contribution to the three largest FCs ($\alpha_1^1$, $\beta_1^1$, and $\alpha_1^2$) is from the pair potential.  Specifically, for these three FCs the embedding-energy contribution is less than 20\% of that from the pair potential, and in most cases the embedding-energy contribution is significantly less.  For Au the situation is slightly more complicated.  For all three Au models the two largest FCs ($\alpha_1^1$ and  $\beta_3^1$) are mainly due to the pair-potential interaction.   However, for remaining first-shell FC ($\alpha_2^1$) and two second-shell FCs ($\alpha_1^2$ and $\alpha_2^2$) the embedding-energy contribution is significantly larger than that from the pair interaction. 

\section{Summary}

In this paper we have studied vibrational dynamics within the EAM formalism.  First, we have derived equations for the dynamical matrix that can be used to model bulk and surface vibrations in materials with multiple atoms per unit cell.  Second, we have simplified these equations to equations that are valid for looking at vibrations in cubic materials with a single atom basis, such as bcc and fcc metals.  Third, we have explored the relationship between the EAM formalism and BvK FCs in bcc and fcc materials.  Lastly, using K, Fe, and Au as examples, we have investigated the relative importance of the various force constants in the ability of an EAM model to predict vibrational dispersion curves.

Our results suggest that one might profitably use BvK FCs as direct inputs when building EAM models.  Typically these force constants are indirectly involved in EAM model construction via the use of elastic constants and/or specific phonon frequencies.  Indeed, this is true of all models discussed above.  Models CM and JO for K, H3 and H5 for Fe, and models PR and CM for Au utilize elastic constants, but not phonon frequencies, while the WR model for K, models H1, H2, H4, and H6 for Fe, and the SH model for Au also utilize phonon frequencies.  In general, the models that utilize both elastic constants and phonon frequencies predict the dispersion curves with more accuracy, although this observation is not universal.  Indeed, model H1 for Fe is the least accurate of the six models for Fe.   The potential advantage of directly using BvK FCs as inputs is that the FCs determine the phonon frequencies throughout the Brillouin zone, not just near the zone center and a few other frequencies.  We are currently investigating the direct utilization of BvK FCs in building EAM models.

\section*{References}

\vspace{-0.5cm}

\bibliography{AlkaliMetals}

\end{document}